\documentstyle[12pt,a4]{article}
\begin{document}

\baselineskip 18pt
\begin{titlepage}

\centerline{\Large{\bf Hydrodynamic behaviour of Lattice Boltzmann}}
\centerline{\Large{\bf and Lattice BGK models}}
\medskip \medskip\medskip\medskip
\centerline{ {\bf O. Behrend, R. Harris}\footnote{Permanent address:\\
Department
of Physics and Centre for the Physics of Materials, McGill University,
Rutherford Building, 3600 University Street, Montr\'eal, Qu\'ebec H3A
2T8, Canada }}
\medskip
\centerline{Department of Physics, The University of Edinburgh,}
\centerline{ Mayfield Road, Edinburgh EH9 3JZ, United Kingdom }
\medskip\medskip
\centerline{and}
\medskip\medskip
\centerline{{\bf P. B. Warren }}
\medskip
\centerline{ Unilever Research, Port Sunlight Laboratory,}
\centerline{ Quarry Road East, Bebington, Wirral L63 3JW, United Kingdom }
\medskip\medskip\medskip
\centerline{(August 30, 1993)}
\medskip\medskip\medskip
\medskip
\centerline{\large Abstract}
\medskip
We present a numerical analysis of the validity of classical and
generalized hydrodynamics for Lattice Boltzmann Equation (LBE) and
Lattice BGK methods in two and three dimensions, as a function of the
collision parameters of these models. Our analysis is based on the
wave-number dependence of the evolution operator. Good ranges of
validity are found for BGK models as long as the relaxation time is
chosen smaller than or equal to unity. The additional freedom in the
choice of
collision parameters for LBE models does not seem to give significant
improvement.

\end{titlepage}

Recently, Lattice-gas Automata (LGA) methods have been developed as a
new computational approach to fluid dynamics \cite{Frisch}. Using purely
local Boolean operations to represent particle collisions, they have proved
to be extremely efficient, although, due to the fluctuations inherent in the
method, statistical averaging is necessary in order to extract information.
The Lattice Boltzmann Equation (LBE) \cite{Succi} and Lattice BGK method
\cite{Chen} \cite{Qian}, by using continuous distribution functions, eliminate
this statistical noise and offer significant computational advantages.

It is important to define ways in which these methods can be optimized.
In the past, since one of the main applications has been the study of
flows at high Reynolds number, it has been customary to minimize the
viscosity by tuning the parameters of the collision operators.
However, since applications to low Reynolds number flows are becoming
increasingly important \cite{Rothman} \cite{Ladd}, other criteria might be
used. In particular, it is of interest to define the spatial scale for which
the models reproduce hydrodynamics, and how this scale depends upon
the parameters of the simulation.

This analysis was first considered by Luo et al. \cite{Luo}, and
developed by Grosfils et al. \cite{Grosfils} and Das and Ernst
\cite{Ernst} for the study of LGA's. They show \cite{Ernst} that some of the
simplest LGA models reproduce classical or even generalised hydrodynamics
\cite{Ernst} only over very large spatial scales, and point out that
observations in the literature of ``negative viscosities'' \cite{Neg} can be
traced to those scale effects. Since, however, there have been no similar
investigations of LBE or BGK methods, we present in
this letter a numerical analysis of the validity of hydrodynamics for
these methods in two and three dimensions, the latter being an extension of
the analysis of four-dimensional models based on the FCHC lattice
\cite{Frisch}.

In the LBE with enhanced collisions \cite{Higuera}, the collision operator
is linearized around the equilibrium distribution function to give the kinetic
equation:
\begin{equation}
f_\alpha ({\bf x+c}_{\alpha},t+1) = f_\alpha ({\bf x},t) + \sum_{\beta =
1} ^{b} \Omega_{\alpha
\beta}\, [f_\beta({\bf x},t) - f_\beta^{equil}({\bf x},t) ]
\end{equation}
where $\Omega_{\alpha\beta}$ is the linearized collision operator,
$f_\alpha ({\bf x},t)$ is the occupation number of velocity ${\bf c}_\alpha$ at
node {\bf x} and time t, $f_\alpha^{equil}({\bf x},t)$ is the chosen
equilibrium
distribution function and b the number of velocity directions. By symmetry,
the matrix element $\Omega_{\alpha\beta}$ depends only on the
angle $\theta$ between the directions $\alpha$ and $\beta$, and by
convention is given the value $a_\theta$ \cite{Higuera}. In two
dimensions, for a single speed model based on the hexagonal lattice
\cite{Frisch} (6
velocity directions), there are four possible matrix elements, $a_0$,
$a_{60}$, $a_{120}$ and $a_{180}$ while for the four dimensional FCHC
lattice (24 velocity directions), there are five possible
matrix elements, $a_0$, $a_{60}$, $a_{90}$, $a_{120}$ and $a_{180}$.
The conditions of mass and momentum conservation then reduce the number of
independent matrix elements, leaving two and three such elements
for the two and four dimensional situations respectively.

Linear stability analysis requires that the eigenvalues of $\Omega$ be
contained in the interval ($-$2,0). This places further restrictions on the
matrix elements, which can be made explicit by writing the non-zero
eigenvalues as \cite{Higuera}:
\begin{eqnarray*}
\lambda = 6(a_0 + a_{60})\\
\sigma = -6(a_0 + 2a_{60})
\end{eqnarray*}
in two dimensions, with multiplicities 2 and 1 respectively, and
\begin{eqnarray*}
\lambda = a_0 - 2a_{90} + a_{180}\\
\sigma = 3(a_0 - a_{180})/2\\
\gamma = 3(a_0 + 6a_{90} + a_{180})/2
\end{eqnarray*}
in four dimensions, with multiplicities 9, 8 and 2 respectively.
It is also useful to note that the eigenvalue
$\lambda$ is linked to the kinematic viscosity of the fluid by \cite{Higuera}
\begin{displaymath}
\nu = - \frac{c^2}{d+2} \left(\frac{1}{\lambda} + \frac{1}{2}\right)
\end{displaymath}
where d is the dimensionality of the simulation.
The eigenvalues $\sigma$ in two
dimensions and $\sigma$ and $\gamma$ in four dimensions control the
decay of the so-called ghost fields \cite{Succi}, and therefore
a preferred choice for their values is $-$1, forcing a rapid decay of
these unphysical fields.

The BGK model (so-called in analogy with the BGK treatment of the
Boltzmann equation \cite{BGK}) is a further simplification of the
LBE model, whereby
the collision part of the kinetic equation is parameterised
by a single relaxation parameter $\tau$ such that \cite{Chen} \cite{Qian}:
\begin{equation}
f_\alpha ({\bf x+c}_{\alpha},t+1) = f_\alpha ({\bf x},t) -
\tau^{-1}\,\cdot\, [f_\alpha({\bf x},t) - f_\alpha^{equil}({\bf x},t)].
\end{equation}
The relaxation parameter $\tau$ (which linear stability requires to be larger
than 1/2, $1/2<\tau<1$ being called subrelaxation and $\tau>1$ over-relaxation)
is linked to the kinematic viscosity by the
relation \cite{Chen}
\begin{displaymath}
\nu = \frac{c^2 (2\tau -1)}{2(d+2)}.
\end{displaymath}
The value $\tau=1$ plays the same role as $\sigma = \gamma =-1$ in the LBE
model since it relaxes the distribution function $f_\alpha ({\bf x},t)$ to
its equilibrium $f_\alpha^{equil}({\bf x},t)$ in a single time step.

To analyse the hydrodynamic behaviour of these models, we use
the properties of the evolution operator  $H({\bf k})$ \cite{Luo}.
Thus, if  $\phi_{\alpha}({\bf x},t)$ is the deviation of $f_{\alpha}({\bf
x},t)$ from the equilibrium distribution,
\begin{displaymath}
\phi_{\alpha}({\bf x},t) = f_{\alpha}({\bf x},t) - f_\alpha^{equil}({\bf x},t),
\end{displaymath}
$H({\bf k})$ can be defined by Fourier
transforming the kinetic equation (1) such that \cite{Luo}:
\begin{displaymath}
|{\bf \phi}({\bf k},t+1)> = H({\bf k})\,|{\bf \phi}({\bf k},t)>.
\end{displaymath}
Here, $<{\bf c}_\alpha|{\bf \phi}({\bf k},t)> = \phi_\alpha ({\bf k},t)$,
$H({\bf k})=D({\bf k}) H(0)$, and $H(0) = I +\Omega$, where  $I$ is the unit
matrix, and the displacement operator $D({\bf k})$ is the diagonal matrix
diag$[exp(-i {\bf k} \cdot {\bf c}_1), exp(-i
{\bf k} \cdot {\bf c}_2), \cdots, exp(-i {\bf k} \cdot {\bf c}_b)]$.
The eigenvalues of $H({\bf k})$, defined from
\begin{displaymath}
H({\bf k})|\psi_\lambda ({\bf k})> = e^{z_\lambda({\bf k})}
|\psi_\lambda ({\bf k})>
\end{displaymath}
then give information about the transport coefficients corresponding to the
collision matrix $\Omega$.

In the long-wavelength regime ($k \rightarrow 0$, where $k=|{\bf k}|$), two
types of modes exist: hydrodynamic modes, related to the conservation laws,
with $Re[z_\lambda({\bf k})]\sim O(k^2)$, and rapidly decaying kinetic
modes, with $Re[z_\lambda({\bf k})] < 0$, without any physical
significance. Transport coefficients are related to the hydrodynamic
modes \cite{Ernst}. In a model without explicit energy conservation in two
(four) dimensions three (five) such modes exist, two (two) propagating but
damped
sound modes ($\lambda = \pm$) and one (three) diffusive shear modes ($\lambda
= \perp$), with $Im[z_\perp({\bf k})] = 0$. The real part
$Re[z_\lambda({\bf k})]$ represents damping, and if the imaginary part
$Im[z_\lambda({\bf k})]= \pm c_s({\bf k})k$ is nonvanishing, the mode
propagates with speed $c_s({\bf k})$ \cite{Ernst}. The wave-vector
dependent kinematic shear viscosity is defined as \cite{Ernst}
\begin{displaymath}
\nu({\bf k}) \equiv - z_\perp({\bf k})/k^2,
\end{displaymath}
while the sound damping constant is defined as
\begin{displaymath}
\Gamma({\bf k}) \equiv -Re[z_\pm({\bf k})]/k^2.
\end{displaymath}

In classical hydrodynamics ($k \rightarrow 0$), the transport coefficients
are $k$-independent by definition. However, when this situation does not hold,
but the hydrodynamic modes are still clearly separated from
the kinetic modes, one can speak of a {\it generalized hydrodynamic}
regime \cite{Ernst}, with transport coefficients which are slowly varying
functions of k. In lattice-based models, the transport coefficients might
also depend on the direction of the wave vector, {\bf \^{k}}, reflecting
anisotropies due to the symmetry of the lattice. By computing the transport
coefficients through the
spectral analysis of the evolution operator, and looking at
their {\bf k}-dependence, one can judge the range of validity of
the classical and the generalized hydrodynamic regime.
In a previous analysis \cite{Ernst}, for the simplest Lattice-gas FHP-I
model with a density of $\rho=1.8$, generalized hydrodynamics were
shown to be valid up to $k\simeq0.4$ for certain directions of {\bf k}
(Figure 3 of reference \cite{Ernst}).

In order to allow for an analysis of $H({\bf k})$ for the Lattice BGK model,
we require  an effective collision matrix. It is easy to verify that, with
\begin{displaymath}
\Omega_{\alpha \beta} = - \frac{1}{\tau} [\delta_{\alpha \beta} -
\frac{1}{b} - \frac{d}{b c^2} \, {\bf c}_\alpha \cdot {\bf c}_\beta],
\end{displaymath}
Eq. (1) reduces to Eq. (2), conservation of both mass and momentum is
satisfied and all the non-zero eigenvalues of $\Omega$ are equal to
$- 1/\tau$. This matrix can therefore be employed for the analysis.

Figure 1(a) shows the real part of a typical spectrum obtained for the
two-dimensional BGK model on a hexagonal lattice with $\tau=3/4$ and {\bf k}
along the {\bf \^{x}} direction (parallel to a lattice vector). One can
distinguish the hydrodynamic (shear and sound) as
well as the kinetic modes ($Re[z_\lambda({\bf 0})] = \ln{|1 - 1/\tau|}$).
Mixing of the two kinds of modes happens at $k\simeq3.0$. Figure 1(b) displays
$-Re[z_\lambda({\bf k})] / k^2$ for the hydrodynamic modes of the same
model, and from this figure
we conclude that classical hydrodynamics is valid up to $k\simeq1.2$
and generalized hydrodynamics up to $k\simeq2.1$. This conclusion is
supported by Figure 1(c) displaying $c_s({\bf k})=\pm Im[z_\lambda({\bf
k})]/k$.
With {\bf k} along the {\bf \^{y}}-direction, the ranges are 1.5 and 2.3
respectively. We have studied the same model for values of $\tau$
ranging from
0.55 to 1.5, and find that the range of classical and generalized
hydrodynamics, given by the behaviour of the real and imaginary parts of
$z_\lambda({\bf k})$, is essentially the same as for the example given above
(i.e. for $\tau=0.75$) as long as $\tau\leq1$. For $\tau>1$, the range
rapidly decreases so that, for instance, at $\tau = 4/3$, classical
hydrodynamics is only valid up to $k\simeq0.3$, and generalized
hydrodynamics up to $k\simeq 1.3$, for {\bf k} along the {\bf
\^{x}}-direction.

Since the two-dimensional LBE model allows for an adjustment of {\it two}
independent parameters,  one might expect a greater scope for
tuning the behaviour of $z_\lambda({\bf k})$. However,
although the quadratic behaviour of
$-Re[z_\lambda({\bf k})]$ for hydrodynamic modes can extend over a greater
range, we find that correspondingly, linear behaviour of
$Im[z_\lambda({\bf k})]$ is found over a smaller range: the overall range
of validity of generalized hydrodynamics is scarcely
improved compared to the Lattice BGK model.
Our ``optimum range'' is very comparable with that for
multispeed FHP Lattice-gas models \cite{Ernst}, where the best results
were obtained for the 7-bit FHP-III model.

Analysis for the four-dimensional LBE
and Lattice BGK models based on the FCHC lattice proceeds similarly. Two and
three dimensional data can be obtained from these four dimensional
models by projecting the lattice onto two or three dimensions and defining
a reduced collision matrix \cite{Succi}. Our numerical studies show
that the spectral behaviour of $H({\bf k})$ constructed with the full
matrix or the reduced matrices is identical, at least for the physically
important hydrodynamic modes. We therefore present results only for the full
four-dimensional FCHC lattice.

For the BGK model, we have again used values of $\tau$ ranging from
0.55 to 1.5. Our findings are that classical hydrodynamics is valid up to
$k\simeq1.0$ and generalized hydrodynamics up to $k\simeq2.0$, independent
of $\tau$, as long
as $\tau\leq1$. As in two dimensions, the ranges rapidly
decrease for $\tau>1$, with, at $\tau = 4/3$, classical hydrodynamics
being only valid up to $k\simeq0.3$, and generalized hydrodynamics up to
$k\simeq 1.3$.
These ranges, although smaller or comparable to those for the two-dimensional
models based on the hexagonal lattice, are still considerable: they
suggests that generalized hydrodynamics is valid down to a spatial scale
of around three lattice spacings.

The four dimensional LBE model allows for an adjustment of {\it three}
independent parameters. As in the two-dimensional case,
the range of validity of generalized hydrodynamics is hardly changed
compared to the Lattice BGK model, but the quadratic behaviour of
$-Re[z_\lambda({\bf k})]$ for the hydrodynamic modes can be tuned and
extended up to $k\simeq1.5$
(for example with $\lambda=-3/2$, $\sigma=\gamma=-1$ as in Fig.2). In
general terms, this seems to correspond to the existence of kinetic
modes which, at {\bf k}=0, are less clearly separated from the
hydrodynamic regime.

We thus conclude that the computational advantages of the Lattice BGK
algorithm are complemented by a significant range of validity for
classical and generalized hydrodynamics. In two dimensions, using a
single speed model on a hexagonal lattice, the range is as good as that
for LGA models. For two- and three-dimensional models based on the FCHC
lattice, the range is as good or better than that for more general LBE
methods. The additional parameters available in the LBE method
seem to give no further advantage. We expect that future applications of
the Lattice BGK algorithm will exploit this situation.

O.B. acknowledges financial support through a Foreign Office - Glaxo joint
scholarship. R.H. acknowledges financial support from the NSERC of
Canada and le Fonds pour la Formation des Chercheurs et l'Aide \`a la
Recherche de la Province du Qu\'ebec.

\newpage
\begin{Large}
\noindent {\bf Captions}
\end{Large}
\medskip\medskip\medskip
\newline
FIG.1.  (a) Real part of the spectrum for the two-dimensional BGK model
with $\tau=3/4$ and {\bf k} $\parallel$ {\bf \^{x}}. The upper of the two
hydrodynamic modes ($-Re[z_\lambda({\bf k})]\sim O(k^2$) as
$k\rightarrow0$) is the diffusive shear mode ($\lambda = \perp$), while
the lower curve corresponds to the propagating sound modes ($\lambda =
\pm$). (b) The corresponding plot of $-Re[z_\lambda({\bf k})]/k^2$ for the
hydrodynamic modes. The upper curve corresponds to the kinematic viscosity
$\nu({\bf k})$, while the lower curve is the sound damping constant
$\Gamma({\bf k})$. (c) The sound velocity $c_s({\bf k}) =
\pm Im[z_\pm({\bf k})]/k$ for the same model.
\medskip\medskip
\newline
\noindent FIG. 2. (a) Viscosity $\nu({\bf k})$ (upper continuous curve)
and sound
damping constant $\Gamma({\bf k})$ (lower continuous curve) for a
four-dimensional LBE model,
with $\lambda=-3/2, \sigma=\gamma=-1$ and {\bf k} $\parallel$ {\bf
\^{x}}. The dashed lines correspond to kinetic modes. (b) The sound
velocity $c_s({\bf k}) = \pm Im[z_\pm({\bf k})]/k$ for the same model.


\newpage
\begin{thebibliography}{00}
\bibitem{Frisch}
U. Frisch, D. d'Humi\`eres, B. Hasslacher, P. Lallemand, Y. Pomeau and
J.-P. Rivet, Complex Syst. {\bf 1}, 649 (1987) [reprinted in {\it
Lattice Gas Methods for Partial Differential Equations}, edited by G.
Doolen (Addison-Wesley, Singapore, 1990)]
\bibitem{Succi}
R. Benzi, S. Succi and M. Vergassola, Phys. Rep. {\bf 222} 147 (1993)
\bibitem{Chen}
S. Chen, Z. Wang, X. Shan and G. D. Doolen, J. Stat. Phys {\bf 68} 379
(1992)
\bibitem{Qian}
Y.H. Qian, D. d'Humi\`eres and P. Lallemand, Europhys. Lett. {\bf 17}
479 (1992)
\bibitem{Rothman}
A.K. Gunstensen, D.H. Rothman, S. Zaleski and G. Zanetti, Phys. Rev {\bf
A 48} 4320 (1991)
\bibitem{Ladd}
A.J.C. Ladd, Phys. Rev. Lett. {\bf 70} 1339 (1993)
\bibitem{Luo}
L.-S. Luo, H. Chen, S. Chen, G.D. Doolen and Y.-C. Lee, Phys. Rev. {\bf
A 43} 7097 (1991)
\bibitem{Grosfils}
P. Grosfils, J.-P. Boon, R. Brito and M.H. Ernst, to appear in Phys.
Rev. E
\bibitem{Ernst}
S.P. Das, H.J. Bussemaker and M.H. Ernst, Phys. Rev. {\bf E 48} 245
(1993)
\bibitem{Neg}
D. d'Humi\`eres and P. Lallemand, Complex Syst. {\bf 1} 599 (1987)
\bibitem{Higuera}
F.J. Higuera, S. Succi and R. Benzi, Europhys. Lett. {\bf 9} 345 (1989)
\bibitem{BGK}
P. Bhatnagar, E.P. Gross and M.K. Krook, Phys. Rev. {\bf 94} 511 (1954)
\end{thebibliography}
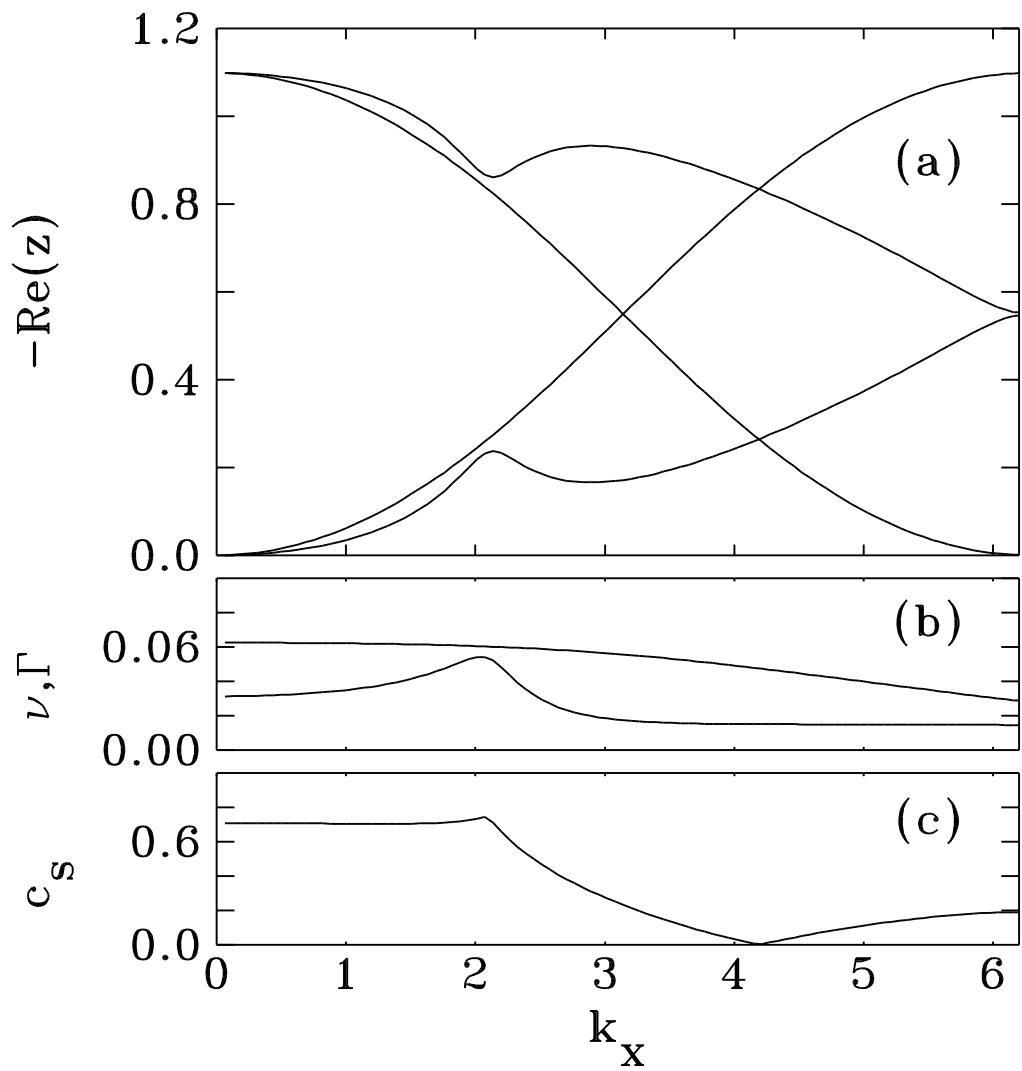
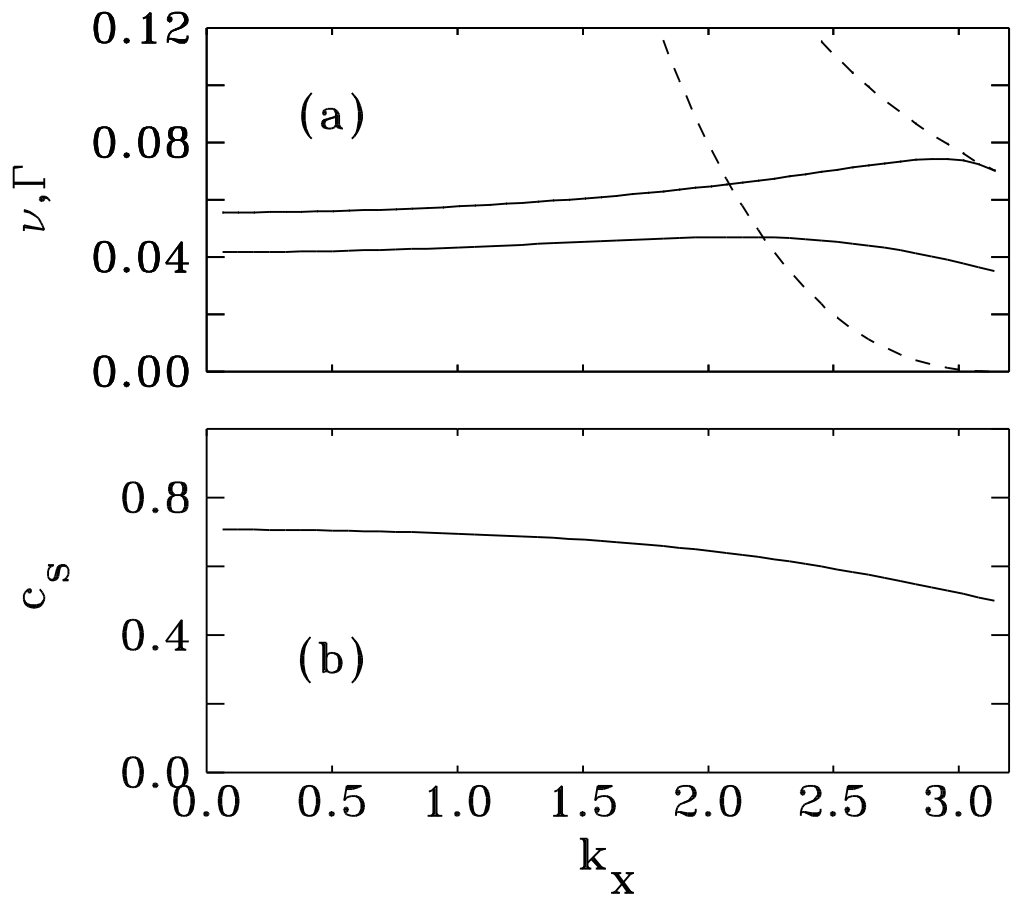
\end{document}